\documentclass[fleqn,usenatbib]{mnras}

\usepackage{newtxtext,newtxmath}

\usepackage[T1]{fontenc}

\DeclareRobustCommand{\VAN}[3]{#2}
\let\VANthebibliography\thebibliography
\def\thebibliography{\DeclareRobustCommand{\VAN}[3]{##3}\VANthebibliography}

\usepackage{graphicx}	
\usepackage{amsmath}	

\title[Apsidal precession in kima]{Improving circumbinary planet detections by fitting their binary's apsidal precession}

\author[T. A. Baycroft et al.]{
Thomas A. Baycroft,$^{1}$\thanks{E-mail: txb187@bham.ac.uk}
Amaury H.M.J. Triaud,$^{1}$
Jo\~ao Faria,$^{2,3}$
Alexandre C.M. Correia,$^{4,5}$
\newauthor
and Matthew R. Standing$^{6}$
\\
$^{1}$School of Physics and Astronomy, University of Birmingham, Edgbaston, Birmingham B15 2TT, UK\\
$^{2}$Instituto de Astrof\'iısica e Ci\^encias do Espa\c co, Universidade do Porto, CAUP, Rua das Estrelas, 4150-762 Porto, Portugal\\
$^{3}$ Departamento de F\'isica e Astronomia, Faculdade de Ci\^encias, Universidade do Porto, Rua do Campo Alegre, 4169-007 Porto, Portugal\\
$^{4}$CFisUC, Departamento de F\'isica, Universidade de Coimbra, 3004-516 Coimbra, Portugal\\
$^{5}$IMCCE, UMR8028 CNRS, Observatoire de Paris, PSL Universit\'e, 77 av. Denfert-Rochereau, 75014 Paris, France\\
$^{6}$School of Physical Sciences, The Open University, Milton Keynes, MK7 6AA, UK
}

\date{Accepted 2023 February 22. Received 2023 January 26; in original form 2022 November 21}

\pubyear{2022}

\begin{document}
\label{firstpage}
\pagerange{\pageref{firstpage}--\pageref{lastpage}}
\maketitle

\begin{abstract}
Apsidal precession in stellar binaries is the main non-Keplerian dynamical effect impacting the radial-velocities of a binary star system. Its presence can notably hide the presence of orbiting circumbinary planets because many fitting algorithms assume perfectly Keplerian motion. To first order, apsidal precession ($\dot{\omega}$) can be accounted for by adding a linear term to the usual Keplerian model. We include apsidal precession in the {\tt kima} package, an orbital fitter designed to detect and characterise planets from radial velocity data. In this paper, we detail this and other additions to {\tt kima} that improve fitting for stellar binaries and circumbinary planets including corrections from general relativity. We then demonstrate that fitting for $\dot{\omega}$ can improve the detection sensitivity to circumbinary exoplanets by up to an order of magnitude in some circumstances, particularly in the case of multi-planetary systems. In addition, we apply the algorithm to several real systems, producing a new measurement of aspidal precession in KOI-126 (a tight triple system), and a detection of $\dot{\omega}$ in the Kepler-16 circumbinary system. Although apsidal precession is detected for Kepler-16, it does not have a large effect on the detection limit or the planetary parameters. We also derive an expression for the precession an outer planet would induce on the inner binary and compare the value this predicts with the one we detect.
\end{abstract}




\section{Introduction}

Exoplanets exhibit a range of configurations much vaster than is present within the solar system. Nearly three decades of discoveries have revealed that most known exoplanets are not analogous to any solar system planets \citep[e.g.][]{winn_occurrence_2015}. This applies to individual planets being different such as Hot Jupiters \citep[e.g.][]{dawson_origins_2018} or planets with extreme eccentricities \citep[e.g.][]{angelo_kepler-1656bs_2022}, but this can also apply to entire planetary systems having more exotic configurations, such as TRAPPIST-1 a multi-planetary resonant chain orbiting a late M-dwarf \citep{gillon_temperate_2016,gillon_seven_2017}. One such type of exotic planetary systems are the circumbinary exoplanets that orbit about both stars of a tight stellar binary \citep{schneider_occultations_1994,doyle_kepler-16_2011}.

To date there have been only 15 fully confirmed circumbinary planets\footnote{Circumbinary planets orbiting stellar remnants are claimed using eclipse timings however there are doubts about their existence and as such we do not consider them as fully confirmed}. All but one have transited at least one of the two stars, and were first detected from space with {\it Kepler} \citep[e.g.][]{doyle_kepler-16_2011, orosz_kepler-47_2012,kostov_kepler-1647b_2016} and with {\it TESS} \citep{kostov_toi-1338_2020,kostov_tic_2021}. The pace of detections is slow, since the two most common exoplanet detection techniques (transit and radial velocity) have so far both been hamstrung, each with their own issues. Circumbinary planets will generally have longer periods than planets around single stars because they need to orbit outside of an instability region produced by the binary stars' motion \citep{dvorak_stability_1989,holman_long-term_1999,doolin_dynamics_2011}. Because of this extra distance, circumbinary planets are geometrically less likely to produce transits than planets orbiting single stars. However, for similarly distant planets, nodal precession makes circumbinary planets more likely to create transits \citep{martin_circumbinary_2015}, even if transits do not happen at every planetary orbit \citep[e.g.][]{schneider_occultations_1994,martin_planets_2014}. 

For the radial velocity method, interference between the spectra of both components of the binary star makes it harder to obtain precise radial velocity measurements \citep[e.g.][]{konacki_radial_2009}. The latter issue can be circumvented by observing single-lined binaries \citep{konacki_high-precision_2010,martin_bebop_2019}. This is the observing strategy employed by the BEBOP survey. BEBOP (Binaries Escorted By Orbiting Planets) has been collecting radial velocities on eclipsing single-lined binaries for over four years and has demonstrated that it can detect circumbinary planets, notably by having independently detected Kepler-16b \citep{triaud_bebop_2022}. There is also the first circumbinary planet discovered in radial velocities BEBOP-1c (Standing et al. submitted).

One significant advantage of the radial velocity method over the transit method is that radial velocities probe the full orbit, instead of just the inferior conjunction. This leads to good precision on the eccentricity \(e\) \citep[sometimes down to {\(10^{-4}\)};][]{triaud_eblm_2017} and the argument of periastron \(\omega\) of the binary orbit. An issue this raises is that variation in \(e\) and \(\omega\) will cause problems when fitting a static Keplerian orbit, a point raised in \citet{konacki_high-precision_2010,sybilski_non-keplerian_2013}. One such variation is apsidal precession of the binary: an evolution of \(\omega\) with time, denoted by \(\dot{\omega}\). This can be caused by relativistic effects, tidal effects, or --most excitingly to exoplanet hunters-- planetary perturbations \citep{correia_tidal_2013}. Because the BEBOP survey has collected data over several years, the scatter caused by this precession is slowly starting to exceed the RMS scatter of the residuals on some systems \citep{standing_bebop_2022}. Accounting for this effect would improve the accuracy of the fits for the binary orbit, and in turn improve our ability to detect planets, and the precision on their physical and orbital parameters. As explored in \citet{standing_bebop_2022}, in most cases the radial-velocity signal of a single planet would be detected well before a non-zero \(\dot{\omega}\) is significantly detected. In this paper we show that, in some multiplanetary configurations, low-amplitude planetary signals can be hidden by the precession induced by another, heavier planet.

In addition, measuring the apsidal precession rate adds new information to our knowledge of the system. Usually, it is not possible to measure orbital inclinations from radial velocities alone. However, the binary's precession rate due to an external perturber is dependent on the mutual inclination between the binary's orbital plane and the perturber's \citep[e.g.][]{correia_tidal_2013,correia_secular_2016}. In this work, we derive an equation to calculate the apsidal precession that a third body induces on an inner binary pair which can be used to calculate the mutual inclination, this can be found in Appendix~(\ref{sec:appA}). In the case of BEBOP, where all binaries are known to eclipse, an upper bound on the mutual inclination directly translates into an upper bound on the orbital inclination of the planet, meaning those radial velocity data can be used to obtain not just a minimum mass $m_{\rm p} \sin i_{\rm p}$, but also a maximum mass.
Finally, for close binaries, measuring the precession rate also provides information on the stars' internal structure \citep{claret_apsidal-motion_2010}.

In this paper, we first we describe a binaries-specific radial velocity model applied in the {\tt kima} package in Sect.~\ref{sec:method}. This new model (which we will occasionally refer to as {\tt kima-binaries} when comparing it with the old model) moves beyond fitting pure Keplerian orbits \citep[as done in][]{faria_kima_2018}, by including an apsidal precession parameter to the fitted model. The section details those changes and describes the inclusion of other tidal and relativistic effects that are known to affect orbital solutions. In Section~\ref{sec:tests}, the new model is used on both simulated and observed data. The ability to accurately recover the apsidal precession rate is demonstrated, and we show how fitting for apsidal precession can improve a survey's sensitivity to circumbinary planets by producing Bayesian detection limits. Finally, in Section~\ref{sec:results} we present a detection of the precession rate for Kepler-16, and conclude in Section~\ref{sec:conc}.

\section{A binary update to {\tt kima}}\label{sec:method}

In this section we present an update to {\tt kima}, developing a binary-specific radial velocity model. This model accounts for various factors that are generally ignored when looking at radial velocities for a single star but recommended when seeking to detect circumbinary planet signals \citep{konacki_radial_2009,sybilski_non-keplerian_2013}. The new model includes tidal and relativistic effects as well as, most notably, apsidal precession of the binary's orbit. The new model is also given the capability to fit double-lined binary data.

{\tt kima} is an orbital fitting algorithm which makes use of diffusive nested sampling \citep[DNest;][]{brewer_diffusive_2011} to sample the posterior distribution for the model parameters. It allows for the number of Keplerian signals being fit to vary freely which is advantageous for Bayesian model comparison. There is a so called "known-object" mode where separate priors can be defined for certain already known signals while allowing to search for further signals freely; this model is ideal to apply to circumbinary systems. As will be discussed later, this method of sampling allows for an efficient method of calculating detection limits.
\subsection{Adding precession to {\tt kima}}
\subsubsection{A linear approximation}
As a first order approximation, we add a linear precession parameter, \(\dot{\omega}\) to {\tt kima}. This parameter is free during a fit and its posterior is estimated. We take the usual equation for the radial velocity of a Keplerian orbit \citep[e.g][]{murray_keplerian_2010}:
\begin{equation}
    V = K(\cos(f+\omega) + e\cos(\omega)) + \gamma,
\end{equation}
with \(\omega\) now being time dependent\footnote{We neglect terms that are order \(\mathcal{O}(t-t_0)^2\)}:
\begin{equation}
    \omega(t) = \omega_0 + \dot{\omega}(t-t_0).
\end{equation}
\(K\) is the semi-amplitude of the radial velocity signal, \(f\) the true anomaly, \(e\) and \(\omega\) the eccentricity and argument of pericentre for the orbit, \(t_0\) is some reference time, and $\gamma$ is the mean velocity of the system (which can be affected by the zero-point calibration of an instrument, but does not impact other parameters). In our case, we use the mean of the times of observation for \(t_0\).

\subsubsection{Period correction}\label{sec:period}
The period of an orbit as a single value is, in any realistic scenario, not completely well defined. Various angles associated with an orbit will vary in time, such as the argument of pericentre \(\omega\) or the mean anomaly \(M\).  Different combinations of these variations could all be called periods. We consider two of these defined as in the Eqs.~(\ref{eq:Pobs}) and (\ref{eq:Pano}). We denote these as {\it observational period}, \(P_{\rm obs}\) which is the time taken peak-to-peak in radial velocity, and  the {\it anomalistic period}\footnote{This nomenclature is often used to refer to the time between two consecutive pericentre passages in precessing systems \citep[e.g.][]{rosu_apsidal_2020,borkovits_triply_2021}}, \(P_{\rm ano}\), which is the time between consecutive pericentre passages.

\begin{equation}\label{eq:Pobs}
    \frac{2\pi}{P_{\rm obs}} \approx \dot{\omega}+\dot{M},
\end{equation}
\begin{equation}\label{eq:Pano}
    \frac{2\pi}{P_{\rm ano}} \approx \dot{M}.
\end{equation}
\(P_{\rm obs}\) is the period that is usually referred to by observational astronomers and can be precisely measured from time between transits or eclipses. In this work we set priors on the binary period based on eclipses so want to use \(P_{\rm obs}\) for this. When including \(\dot{\omega}\) into radial velocity fits we want to use \(P_{\rm ano}\) as the period parameter to avoid the expected correlation between \(P_{\rm obs}\) and \(\dot{\omega}\). Hence we need to be able to convert from one to the other. To do this we combine the two equations to get
\begin{equation}
    \frac{2\pi}{P_{\rm obs}} = \dot{\omega} +  \frac{2\pi}{P_{\rm ano}} ,
\end{equation}
and hence, neglecting terms of order \(\mathcal{O}(\dot{\omega}P)^2\),\

\begin{align}
    &P_{\rm ano} = \frac{P_{\rm obs}}{\left(1 - \frac{\dot{\omega}P_{\rm obs}}{2\pi}\right)} \approx P_{\rm obs}\left(1 + \frac{\dot{\omega}P_{\rm obs}}{2\pi}\right),\\
    &P_{\rm obs}  = \frac{P_{\rm ano}}{\left(1 + \frac{\dot{\omega}P_{\rm ano}}{2\pi}\right)} \approx P_{\rm ano}\left(1 - \frac{\dot{\omega}P_{\rm ano}}{2\pi}\right).
\end{align}
Our model fits for \(P_{\rm obs}\) as a parameter (i.e. the period prior is for \(P_{\rm obs}\) as is the output posterior distribution), but the model internally converts this to \(P_{\rm ano}\).

\subsection{Other additions to the binaries model}
Here we describe the other additions made to the binary model on top of the apsidal precession described above, namely, we add relativistic and tidal corrections, and give the ability to fit the radial velocities for a double-lined binary.

\subsubsection{Relativistic and tidal corrections}
We include relativistic corrections for binary orbits, the main ones being light-travel time and transverse doppler \citep[LT,TD;][]{sybilski_non-keplerian_2013} and gravitational redshift \citep[GR;][]{zucker_spectroscopic_2007}\footnote{\citet{sybilski_non-keplerian_2013} also have an equation for the gravitational redshift, but it contains errors, hence we use the equation from \citet{zucker_spectroscopic_2007}}:
\begin{align}
    \Delta V_{\rm LT} &= \frac{K_1^2}{c}\sin^2(f+\omega)(1+e\cos{f}),\\
    \Delta V_{\rm TD} &= \frac{K_1^2}{c\sin^2{i}}\left(1+e\cos{f}-\frac{1-e^2}{2}\right),\\
    \Delta V_{\rm GR} &= \frac{K_1(K_1+K_2)}{c\sin^2{i}}(1+e\cos{f}),
\end{align}
where \(e\), \(f\), \(\omega\), and \(i\) are respectively the eccentricity, true anomaly, argument of pericentre, and inclination of the binary orbit relative to the plane of the sky; \(K_1\) and \(K_2\) are the semi-amplitudes of the primary and secondary, respectively, and \(c\) is the speed of light.

The tidal effect is calculated as in \citet{arras_radial_2012}, assuming a circular orbit. The equation for the tidally induced radial velocity signal is as follows:
\begin{equation}\label{eq:tide}
    v_{\rm tide} = 1184\,\frac{M_2R_1^4}{M_1(M_1+M_2)}P^{-3}\sin^2i\sin[2(f-\phi_0)]\:\rm{m\,s^{-1}},
\end{equation}
where \(M_1\), \(M_2\), and \(R_1\) are the mass and radius of the primary and secondary (in solar units), \(P\) the orbital period (in days) (we use \(P_{\rm{ano}}\)), \(f\) the true anomaly, and \(\phi_0 = \pi/2 - \omega\) is the observer's reference position. 

These equations are incorporated as an optional feature into the model such that when a binary model is fit, these contributions to the radial velocities can be naturally accounted for. We do not include these effects for the general planet search objects as their effects will be much smaller (by orders of magnitude since these corrections scale with \(M^2\)) and so does not warrant the increase in computation time.

\subsubsection{Adding in a double-lined binary model}
The vast majority of spectroscopic binaries are double-lined \citep[e.g.][]{kovaleva_visual_2016}, and although the detection of circumbinary planets in double-lined system is problematic \citep[as in][]{konacki_radial_2009,konacki_high-precision_2010}, new methods to disentangle both spectral components accurately enough to detect circumbinary planets are being developed (e.g. Lalitha et al. in prep). To prepare for the time when circumbinary planets can be searched for, and be detected in double-lined binaries, we add a feature to {\tt kima} to model such a configuration. As an input, the software requires files containing radial-velocities for each component of the binary. The sets of data are fit simultaneously, each with an independent $\gamma$ parameter to account for differing zero-point calibrations\footnote{Even though one would expect the same $\gamma$ for components observed with the same instrument, this may not be the case \citep{southworth_solar-type_2013}}. In addition, each set has its own {\it jitter} term, added in quadrature to the RV uncertainties to account for any additional sources of white noise. Only one extra common parameter is fit, the mass ratio \(q\). 

Any given solution consists of a binary orbit, some number of planetary orbits, and polynomial trends up to cubic order. The binary orbit is fit to each dataset with the secondary having the semi-amplitude K scaled by q, and its argument of periastron reversed \(\omega_2 = \omega_1 - \pi\). The planetary orbits are then fit in the same way to each dataset just as {\tt kima} usually does. 

\subsection{Using the model}

The additions in the new binary model can be used in various combinations. The tidal correction and relativistic correction can each be turned on or off, they will then apply to any known objects included. A prior on \(\dot{\omega}\) will need to be given for each known object as well as for general signals. 

One important thing to note is that Eq.~(\ref{eq:tide})  assumes a circular orbit.
We therefore recommend not using the tidal correction for eccentric binaries.
We currently assume an inclination of \(90^\circ\), and therefore we only consider eclipsing systems in this paper. A future update may include the inclination as a free parameter to either attempt to constrain or at the very least marginalise over.

The use of double lined binaries is also included in the options. This requires a dataset (or multiple) with 5 columns: date; RV of primary; uncertainty on primary RV; RV of secondary; uncertainty on secondary RV. 
The primary will therefore automatically be the signal placed in the second column. The mass ratio \(q\) can be larger than 1 (at which point the "secondary" is actually the more massive star), so for example in an almost equal mass case a prior can straddle \(q=1\).

\section{Performances of {\tt kima-binaires}}\label{sec:tests}
We now show tests and applications of the binaries model using data from simulations as well as from real systems. We first show that the model is able to recover consistent values of apsidal precession, and demonstrate the improvement in the fit that ensues. We illustrate this improvement by computing detection limits.

The standard way to perform detection limits is to inject a fine grid of simulated Keplerian signals (often assuming $e=0$) into the data where any planetary signal has been removed, and to measure which signals are recovered by the algorithm \citep[e.g.][]{konacki_radial_2009,konacki_high-precision_2010,mayor_harps_2011,bonfils_harps_2013,rosenthal_california_2021}. Here, instead, we use the posterior distribution of the undetected Keplerian signals to measure the amplitudes which can still be present in the data, as described in \citet{standing_bebop_2022}. 

Briefly, if the analysis indicates there are no planets in a system (circumstellar or circumbinary), we then apply a strict prior on the number of planets by fixing $N_{\rm p} = 1$. One Keplerian signal is assumed to be present and the algorithm is thus forced to return all solutions that are compatible with the data, but not formally detected. We then analyse the posterior samples and compute a limit of $K$ as a function of $P$ that envelops the lower $99\%$ of the samples. Practically speaking, the limit is produced by creating log-linear bins along the $P$ axis. It is best to ensure there are at least at least 1000 samples in each bin. Should a system have a formally detected planet (Bayes factor exceeding 150), that planet is subtracted from the data (the maximum likelihood parameters are used for this), and the detection limit is then computed as above using the residual radial velocities. 

The advantage of using such a method over traditional methods is the ability to sample over all orbital parameters as finely as the algorithm allows (indeed also, all $\gamma$ variables and {\it jitters}, as well as \(e\), \(\omega\), \(\dot{\omega}\), and \(\phi_0\) which are sometimes avoided by the traditional methods). While a traditional insertion/recovery asks the question {\it can these exact signals be recovered?} our method instead asks {\it what is compatible with the data?} A planet below the detection threshold is consistent with the data, and thus, is not formally detected, whereas a planet above the line is inconsistent with the data and would therefore have been detected had it been there.

\subsection{Analytic equation for precession}

In appendix \ref{sec:appA} we derive an analytic equation for the expected precession rate due to an outer perturber (Eq.\,(\ref{eq:wdottransit})), and due to rotational and relativistic effects (Eq.\,(\ref{eq:wdot_corr})). This is done in a similar way to in \citet{correia_tidal_2013} but where that was done in the invariant plane, we do the calculation in the sky plane, which is directly applicable to observations results.

The precession due to a perturber (Eq.\,(\ref{eq:wdottransit})) is under the assumption of both an eclipsing and transiting system, such as Kepler-16. If applying this to a system that does not conform to these assumptions, then Eq.\,(\ref{eq:wdot}) should be used.

\subsection{Testing {\tt kima-binaries} with simulated data}

We begin by testing our ability to recover the apsidal precession using simulated data, showing that we can recover a good measurement of the apsidal precession rate $\dot{\omega}$, and that including it can greatly improve the fit. 

We perform two simulations, both with a primary star of mass \(M = 1\) \(\rm M_{\odot}\), a secondary with \(M = 0.37~\rm M_{\odot}\), \(P = 21.08\) days, and \(e = 0.16\) and a (roughly Jupiter mass) planet with \(M = 0.001\) \(M_{\odot}\), \(P = 134.5\) days, and \(e = 0.01\). The first simulation, SIM1, has just these three bodies, whereas the second simulation, SIM2, has an additional planet with \(M_{\rm pl} = 0.00015~ \rm M_{\odot}\), \(P = 911.2\) days, and \(e = 0\), corresponding to about 3 times the mass of Neptune. We chose these parameters to emulate a typical circumbinary planet: the binary is similar to Kepler-16 in mass-ratio and eccentricity, with a shorter period to increase the amount of precession that will have happened across the time that we "observe". Planet 1 was placed between the 6:1 and 7:1 mean-motion resonances with the binary and planet 2 at a similar period ratio again. Three different masses of planet 2 were tried and we report here the one that had the right mass to be missed without using precession but detected when including it. The decimal places for the periods were chosen randomly to try and avoid integer numbers of days and potential accidental resonances.

Simulations are made using the {\tt rebound} package \citep{rein_rebound_2012}, the integrations used the {\tt IAS15} integrator \citep{rein_ias15_2015}. Radial velocity simulations are taken as the velocity along the line-of-sight within the simulation. The simulation uses the same observational cadence as for Kepler-16 \citep[][]{triaud_bebop_2022}, thus producing a simulated dataset including all Newtonian perturbations. Both simulated datasets are given a Gaussian white noise.
\begin{table}
	\centering
	\caption{For SIM1 we show the parameters from {\tt rebound} \citep{rein_rebound_2012} (note these are keplerian parameters taken from a Newtonian simulation) alongside the fitted parameters both with precession ({\tt kima-binaries}) and without ({\tt kima}). For the planet we state the upper bound on the eccentricity and omit the angle parameters as these are not resolved (close to circular orbit). The 1\(\sigma\) uncertainties are shown as the last few significant digits, all of which are on the same scale as the smallest uncertainty to allow for easy comparison. Goodness-of-fit parameters are also shown to compare the two fits.}
 	\label{tab:Sim1_table}
 	\resizebox{\columnwidth}{!}{
	\begin{tabular}{lcccr}
		\hline
		 & {\tt rebound} &{\tt kima} & {\tt kima-binaries} & units\\
		\hline
		\(P_{\rm{B}}\) & 21.0805330(8492) & 21.0810474(52) & 21.0810395(21) & days\\
		\(M_{\rm{B}}\) & 0.37 & 0.3700886(153) & 0.3700981(57) & \(\mathrm{M_{\odot}}\)\\
		\(K_{\rm{B}}\) & 23415.30(45) & 23419.67(80) & 23420.11(31) & \(\mathrm{ms^{-1}}\)\\
		\(e_{\rm{B}}\) & 0.160106(53) & 0.160114(34) & 0.160096(15) & \\
		\(\omega_{\rm{B}}\) & 4.50018(236) & 4.50061(24) & 4.50016(20) & rad\\
		\(\phi_{0,\rm{B}}\) & 6.27400(23628) & 6.28256(24) & 6.28288(26) & rad\\
		\(\dot{\omega}_{\rm{B}}\) & 308.4(3.7) & 0 & 304.0(15.8) & \(\mathrm{arcsec\,yr^{-1}}\)\\
		& & & & \\
		\(P_{\rm{pl}}\) & 135.084(1.147) & 131.373(120) & 131.392(57) & days\\
		\(M_{\rm{pl}}\) & 1.0476 & 1.076(27) & 1.046(14) & \(\mathrm{M_{J}}\)\\
		\(K_{\rm{pl}}\) & 33.62(09) & 34.81(90) & 33.82(43) & \(\mathrm{ms^{-1}}\)\\
		\(e_{\rm{pl}}\) & 0.0161(79) & <0.050 & <0.033 & \\
		& & & & \\
		RMS & & 6.61 & 3.15 & \(\mathrm{ms^{-1}}\)\\
		Jitter & & 6.27 & 2.04 & \(\mathrm{ms^{-1}}\)\\
		\(\chi_{\nu}^2\) & & 12.35 & 3.15 & \\
		\hline
	\end{tabular}
	}
\end{table}
\begin{figure}
    \includegraphics[width=\columnwidth]{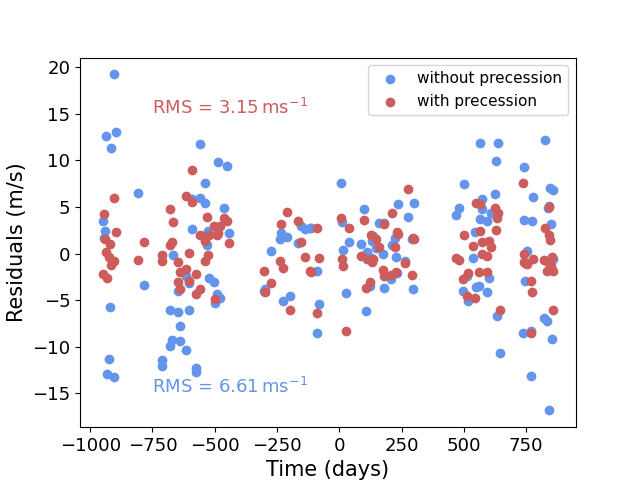}
    \caption{A comparison of radial velocity residuals, after removing the binary's orbital solution, for SIM1 where apsidal precession is included in the {\tt kima-binaries} model (in red) and not included in {\tt kima} (in blue).}
    \label{fig:sim_resid}
\end{figure}
\begin{figure}
    \includegraphics[width=\columnwidth]{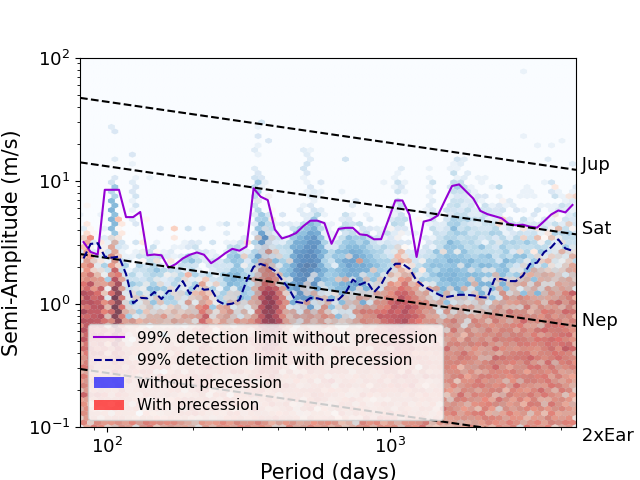}
    \caption{Detection limits for additional Keplerian signals using SIM1. The hexbins represent the density of posterior samples in each run. The purple dashed line and solid blue line are the 99\% detection limits. The black dashed lines show where bodies of various masses would sit on this plot.}
    \label{fig:sim_detlim}
\end{figure}

\subsubsection{Improving scatter and derived parameters}

First, we consider SIM1. From the values of \(\omega\) at each point in the {\tt rebound} simulation, we obtain \(\dot{\omega} = 308.4 \pm 3.7 \,\rm arcsec\, yr^{-1}\) for the binary. Using Eq.\,(\ref{eq:wdottransit}) we get a theoretical value of \(\dot{\omega} = 301.1^{+0.8}_{-1.5}\,\rm arcsec\, yr^{-1}\) A fit using the binaries model results in a posterior estimate of \(\dot{\omega} = 304 \pm 16\) arcsec/yr, which is in agreement with both the simulated and theoretical values. This run is done with the apsidal precession of the binary fit for, but without the relativistic or tidal corrections. The uncertainty in the {\tt rebound} value comes from "sampling" \(\omega\) at various times, calculating \(\dot{\omega}\) from these and then taking its mean and variance. The uncertainty on the theoretical value is propagated in a monte-carlo way from the posterior uncertainty on the binary and planetary parameters. The {\tt kima-binaries} value's uncertainty is defined from the $16^{\rm th}$ to the $84^{\rm th}$ percentiles of the posterior distribution.

Table \ref{tab:Sim1_table} lists the parameters of the binary and planet taken from {\tt rebound} and fit with precession ({\tt kima-binaries}) and without precession ({\tt kima}). The 1\(\sigma\) uncertainty in each measurement is shown in brackets as the last few significant figures, to make comparison easier these are all scaled so that each value on a row is shown to the same number of decimal places. The parameters for {\tt rebound} are read out as the osculating parameters at the times of each datapoint and then the mean and standard deviation of the values are calculated. 

We note that in many cases the fitted values are inconsistent with the {\tt rebound} values, and give a word of warning for using the Keplerian parameters from an n-body fitter such as this. When taking the Keplerian orbital parameters of a body from a {\tt rebound} simulation at a given time, these are taken from the osculating Keplerian orbit which may not be representative of the average orbit. Consider the planet's orbital period, {\tt rebound} effectively gives us an anomalistic period as defined in Sect.~\ref{sec:period}. Because of the perturbed motion this is not the time it will take to actually complete one orbit and we get an observed period a few days shorter. This effect cannot be fit as an apsidal precession of the planet as the orbit is not detectably eccentric.

So a Keplerian (or quasi-Keplerian) fit does not reproduce the osculating Keplerian parameters from a n-body simulation, but it does (to a reasonable accuracy) reproduce the mass. We see in table \ref{tab:Sim1_table} that the mass of the binary is accurate to 3 decimal places (which is more than the precision we usually get on the mass of the primary star anyway) and the mass of the planet is accurately characterised (more so when apsidal precession is take into account).

We can also compare the precision of the two fits, in the sense of how tight a posterior distribution we get for each parameter. In most cases we can see an improvement in precision by about a factor of two.

The reduction in residual scatter can be seen in Figure \ref{fig:sim_resid} where the Root-Mean-Square improves from \({\rm RMS} = 6.61~\rm m\,s^{-1}\) to \(3.15~\rm m\,s^{-1}\). When apsidal precession is not accounted for, if we move further from the reference time \(T_0\) (near the centre of the figure), the fit worsens, giving a characteristic \textit{bow-tie} shape, but if precession is accounted for, the spread in the residuals is reduced. 

The detection limits both with and without precession can be seen in Figure \ref{fig:sim_detlim}. The improvement in detection limit is slightly larger at high periods where the radial-velocity signature of apsidal precession can be confused for a long-term trend. This improvement means the data would allow the detection of another planet signal within this system, almost an order of magnitude lower in mass at orbital periods between \(1,000\) and  \(2,000~\rm days\). Whilst this sounds impressive, this simulation only had a very small amount of extra white noise added to maximise the effect of apsidal precision in order to reveal its importance. In other systems we may expect more marginal improvements (see Sect.~\ref{sec:results}).

\subsubsection{Detecting a hidden planet}
Here, we consider SIM2 and test how many planets are formally detected. To register as an \(n\)-planet detection, the Bayes Factor for the \(n\)-planet solution compared to the \((n-1)\)-planet solution needs to be greater than 150. The sampling in {\tt kima} is trans-dimensional, meaning that solutions with different numbers of planets are all searched simultaneously. Therefore, the Bayes Factor \(BF_{i+1,i}\) comparing the model with \(i+1\) planets to that with \(i\) planets, is the ratio of the number of posterior samples with \(i+1\) planets \(N_{i+1}\) to the number with i planets \(N_i\)
\begin{equation}
    BF_{i+1,i} = \dfrac{N_{i+1}}{N_{i}}
\end{equation} 
Should \(N_i = 0\), the Bayes Factor in Eq.~(12) becomes infinite. This can happen if the BF is larger than the number of effective posterior samples, and would be solved eventually had the sampling continued. In this case we therefore choose to report \(BF_{i+1,i}=N_{i+1}\), effectively setting \(N_i = 1\). More information can be found in \citet{faria_kima_2018,standing_bebop_2022,triaud_bebop_2022}.

Running {\tt kima} on the data from SIM2 without including precession, the outer planet is not formally detected but visible within the posterior as an over-density. With \(BF_{2,1}=12.5<150\), it would be classified as a candidate planet. We can attribute this non-detection to the apsidal precession since when we do fit for the precession, the planet is formally detected with \(BF_{2,1}>6086>150\). The Bayes Factors for each attempted fit are found in Table \ref{tab:BF_table}.
\begin{table}\label{tab:bfs}
	\centering
	\caption{The Bayes Factors for SIM2 (containing two circumbinary planets) comparing models with increasing numbers of planets both for the standard version of {\tt kima} and the new {\tt kima-binaries} version, which includes apsidal precession.}
 	\label{tab:BF_table}
	\begin{tabular}{lcr} 
	\hline
	 & {\tt kima} & {\tt kima-binaries}\\
	 \hline
	 \(BF_{1,0}\)& 538 & 0 \\
	 \(BF_{2,1}\)& 12.5 & 6086 \\
	 \(BF_{3,2}\)& 0.9 & 0.8 \\
	 \hline
    \end{tabular}
\end{table}

\subsection{Testing {\tt kima-binaries} on data from the KOI-126 system}
In this section we test the ability to recover a value for the precession rate consistent with a previous solution from literature, as well as show the improvement in sensitivity to planets that accounting for precession could bring in a highly precessing system.

KOI-126 is a compact triply-eclipsing  hierarchical triple star system. It contains a roughly circular, low-mass, tight binary which is in an eccentric orbit about a more massive tertiary star. The system was first reported in \citet{carter_koi-126_2011}. There are radial velocity data as well as photometry during multiple eclipses. As such, the apsidal precession rate of the tertiary orbit is well measured with a period of $21\,850\,\rm days$ \citep{yenawine_photodynamical_2022} which corresponds to \(\dot{\omega} = 21\,650\,\rm arcsec\,yr^{-1}\). We used 29 radial velocity data from \cite{yenawine_photodynamical_2022}. Our fit with the new binaries model recovers a consistent value for the apsidal precession, with \(\dot{\omega} = 21\,800 \pm 600\,\rm arcsec\,yr^{-1}\). Equivalently this is \(\dot{\omega} = 0.56 \pm 0.009\) degrees per cycle. 

\begin{figure}
    \includegraphics[width=\columnwidth]{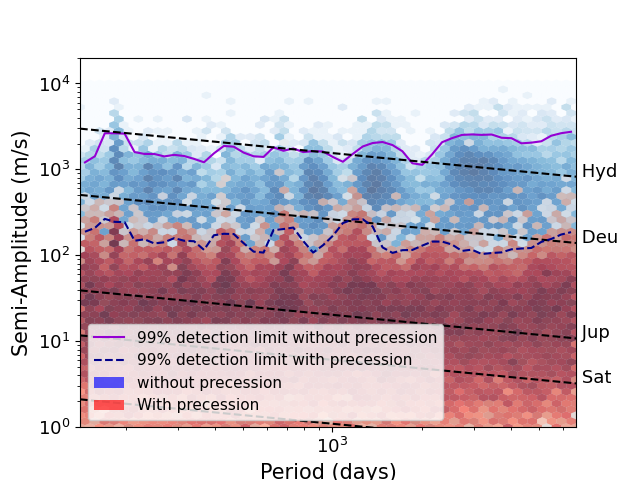}
    \caption{Detection limits for additional signals around KOI-126. The hexbins show the density of posterior samples with the red being those when precession is included in the fit, blue when it is not included. The dashed purple line and solid blue line show the 99\% confidence detection limits. The dashed lines show where bodies of various masses, and where the Deuterium and Hydrogen fusing limits would sit on this plot.}
    \label{fig:KOI126_detlims}
\end{figure}

We run the analysis with apsidal precession fit for, as described in Sect.~\ref{sec:method}, but do not include either the general relativity or the tidal corrections. A more complete analysis would be to use a Newtonian model, rather than Keplerian with added precession, as in \citet{yenawine_photodynamical_2022}, however including the precession improves the \(\chi^2_{\nu}\) from \(640.9\) to \(1.5\). This amply justifies adding an extra parameter, $\dot{\omega}$, to the fit and suggests that a full dynamical model is not necessary with the current precision of the data, hence illustrating the importance of $\dot{\omega}$ since, even in this dynamically complex triple system, the linear apsidal precession removes the majority of the excess noise. The detection limits are shown in Figure \ref{fig:KOI126_detlims} for reference. Here too, including precession improves the detection limit by an order of magnitude in semi-amplitude and removes much of the long-period noise where, as with the simulated data, the precession may be being mildly confused for a long term trend.

We do not use the analytic equation derived in Appendix\,(\ref{sec:appA}) as KOI-126 is in a different orbital configuration where the precessing orbit is the outer (rather than inner) one.

As an interesting note, the orbital periods shown in Figure \ref{fig:KOI126_detlims} would be for putative {\it circumtertiary planets}, of which none are known in Nature. We can nonetheless state there are no stellar or brown dwarf mass companions within \(\sim10^4~\rm days\) of the inner tertiary.

We show the parameters from our fit of KOI-126 in Table \ref{tab:KOI-126}

\section{Application of {\tt kima-binaries} to Kepler-16}\label{sec:results}
\begin{figure}
    \centering
    \includegraphics[width=\columnwidth]{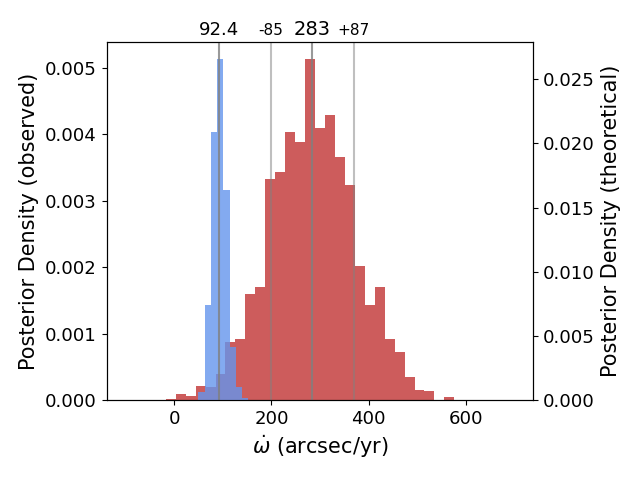}
    \caption{Kepler-16: red: histogram of the density of posterior samples for fitted value of \(\dot{\omega}\) with the median and 1\(\sigma\) values shown in grey. blue: histogram of the density of posterior samples for the theoretically calculated value of \(\dot{\omega}\) with the median value shown in grey. (Note that \(\dot{\omega}\) is not cut at zero, there are in fact posteriors below zero.)}
    \label{fig:K16_wdot}
\end{figure}
The announcement of Kepler-16b marked the first unambiguous detection of a circumbinary planet \citep{doyle_kepler-16_2011}, made thanks to the {\it Kepler} spacecraft \citep{borucki_kepler_2010}. This system is unique in also being the only circumbinary planet independently detected with radial velocity \citep{triaud_bebop_2022}. We re-analyse these radial-velocity data with with our new model, and successfully detect an apsidal precession rate of \(\dot{\omega}_1 = 283^{+87}_{-85}~\rm arcsec\,yr^{-1}\), which is $3.3\sigma$ from 0. Using Eqs.\,(\ref{eq:wdottransit},\,\ref{eq:wdot_corr}) we obtain a value of \(\dot\omega_1 = 92.4^{+14.3}_{-13.8} ~\rm arcsec\,yr^{-1}\), which is 2.2 \(\sigma\) away from the observed value. This theoretical value takes into account the planetary-induced and relativistic precessions (we do not include the rotational and tidal contributions as they would be very small in comparison and parameters like the Love numbers are not very well known).

The theoretical $\dot{\omega}$ is lower than the value that we measure; more data are required to determine how significant this discrepancy is. The difference is likely too important to be accounted for entirely by mutual inclination. An alternative (or additional) explanation could be further undetected planets contributing to the precession rate.

We explore the difference between \(P_{\rm obs}\) and \(P_{\rm ano}\). These values for Kepler-16 are presented in Table \ref{tab:P_table} alongside values published in \citet{triaud_bebop_2022}. The value we get for \(P_{\rm obs}\) is in statistical agreement with the previous publication, however \(P_{\rm ano}\) is \(3.3\sigma\) above this. \(P_{\rm ano}\) is the time between consecutive pericentre passages, the period that should in theory be used to compute physical parameters such as semimajor axis and planet mass, in a Keplerian context. In practice the difference is negligible due to there being a small difference between \(P_{\rm ano}\) and \(P_{\rm obs}\) as well as the uncertainty in the mass of the primary often being dominant. For Kepler-16 the difference in mass using the two periods is \(\approx 2\,\times\,10^{-6}\,\mathrm{M_{\odot}}\). It would take a case with very precise mass and a very high precession rate for this difference to be significant, even for KOI-126 (B+C), the difference is \(\approx 2\,\times\,10^{-4}\,\mathrm{M_{\odot}}\) which is about a fifth of the currently measured uncertainty.

\begin{figure}
    \centering
    \includegraphics[width=\columnwidth]{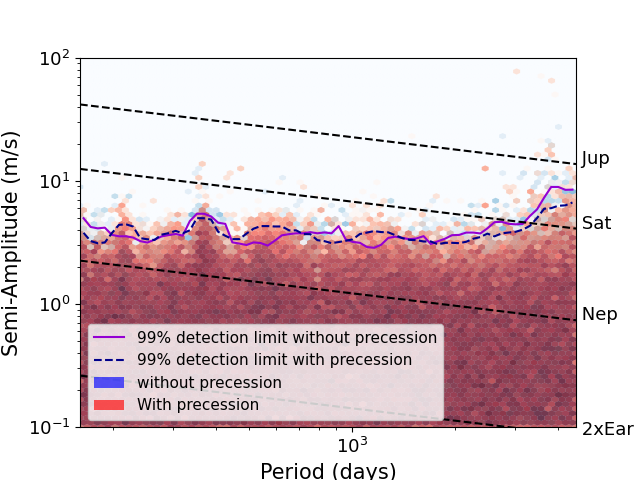}
    \caption{Detection limits for Kepler-16 with and without accounting for precession. The posterior sample for each of these are plotted in the hexbins with the version including precession plotted in red and in front. Since there is no improvement the density plot for the model without precession is hard to see. The dashed purple and solid blue lines show the  99\% confidence detection limits with and without including precession.}
    \label{fig:K16detlims}
\end{figure}

In addition, we produce a detection limit for Kepler-16, comparing the results with and without including \(\dot{\omega}\). The detection limits, plotted in the same way as the previous ones, are shown in Figure \ref{fig:K16detlims}. In this case, as we only get a marginal detection of apsidal precession there is no real improvement in the detection limits.

\begin{table}
    \centering
    \begin{tabular}{l|c}
        \hline
        Type of Period and algorithm used & Value \\
        \hline
         \(P\) ({\tt yorbit})* & 41.077779(54) \\
         \(P\) ({\tt kima})* & 41.077772(51) \\
         \(P_{\rm obs}\) ({\tt kima-binaries}) & 41.077716(55) \\
         \(P_{\rm ano}\) ({\tt kima-binaries}) & 41.078737(305) \\
         \hline
    \end{tabular}
    \caption{Various binary periods for Kepler-16AB. The first two values (*) are taken from \citet{triaud_bebop_2022} using their two different algorithms, the other two are obtained using {\tt kima-binaries} 1\(\sigma\) uncertainties are shown in brackets as the last 2 significant figures (3 in the last case for easy comparison with the others)}
    \label{tab:P_table}
\end{table}

We show the parameters from our fit of Kepler-16 in Table \ref{tab:Kepler-16}.

\section{Conclusions}\label{sec:conc}

We have shown that fitting for the apsidal precession of a binary's orbital parameters improves the radial velocity sensitivity to circumbinary planets. Our conclusions are in line with previous work such as \citet{konacki_high-precision_2010} and \citet{sybilski_non-keplerian_2013}, but extend theirs to a fully Bayesian framework. The improvement in the detection limits can be of up to an order of magnitude in some configurations, but in most cases, improvements are expected to be marginal as in the case of Kepler-16. Accounting for precession can also give improvements in the precision of the parameters recovered from a fit, as well as the potential to uncover planets that were hidden by precession (or instead require less data to detect the same planet). 

We have derived a formula for calculating the theoretical precession induced in a binary (Eq.\,(\ref{eq:wdottransit})) and have discussed the potential use of a measurement of the apsidal precession rate as a way to constrain the mutual inclination of the planetary and binary orbital planes using this formula.

The theoretical and observed values of precession for Kepler-16 are in slight tension, this may be because of undetected planets or some other unknown mechanism.

The longer the baseline of radial velocity observations, the more important it is to account for apsidal precession. As the field progresses, and the number of data from surveys like BEBOP increases, fitting the apsidal precession of the binaries will become vital. To prepare for that time, we have presented an updated version of the {\tt kima} package which is more adapted to fitting radial velocities for single and double-lined binaries. The code is made public on github.

\section*{Acknowledgements}
The authors thank the anonymous reviewer as well as Darin Ragozzine for their useful comments.
This research is in part funded by the European Union's Horizon 2020 research and innovation programme (grants agreements n$^{\circ}$ 803193/BEBOP). A.C. acknowledges support from CFisUC (UIDB/04564/2020 and UIDP/04564/2020), GRAVITY (PTDC/FIS-AST/7002/2020), and ENGAGE~SKA (POCI-01-0145-FEDER-022217), funded by COMPETE 2020 and FCT, Portugal.
MRS acknowledges support from the UK Science and Technology Facilities Council (ST/T000295/1).

\section*{Data Availability}
The radial velocity data for Kepler-16 can be found in \citet{triaud_bebop_2022}.The radial velocity data for KOI-126 can be found in \citet{yenawine_photodynamical_2022}.

The code used for the analysis in this paper can be obtained at https://github.com/j-faria/kima



\bibliographystyle{mnras}
\bibliography{Bibliography}



\appendix

\section{Derivation of the planetary induced precession rate}\label{sec:appA}
In this section we derive the equation for the apsidal precession of an inner orbit (Binary or planet) due to an outer perturber.

The dominating term for the apsidal precession is the planet-binary gravitational interactions.
We take the secular quadrupole Hamiltonian after averaging over the mean anomaly of both orbits is given by \citep[e.g.][]{farago_high-inclination_2010,morais_precession_2012} 
\begin{equation}
    \mathcal{H} = C_2\left[2 - 12e_1^2 - 6(1-e_1^2)\cos^2 i+30e_1^2\cos^2 \alpha \right],
\end{equation}
where 
\begin{equation}
    C_2 = \frac{\mathcal{G}}{16}\frac{m_0m_1}{m_0+m_1}\frac{m_2}{(1-e_2^2)^{3/2}}\frac{a_1^2}{a_2^3},
\end{equation}
and 
\begin{equation}
    \cos i = \sin I_1 \sin I_2 \cos(\Omega_1-\Omega_2)+\cos I_1 \cos I_2,
\end{equation}
\begin{equation}
    \begin{aligned}
    \cos \alpha = &  \sin I_1 \sin \omega_1 \cos I_2 
    - \sin I_2 \cos \omega_1 \sin(\Omega_1 - \Omega_2) \\
    &-\sin I_2 \cos I_1 \sin \omega_1 \cos(\Omega_1 - \Omega_2) .
    \end{aligned}
\end{equation}
In these equations, \(a\), \(e\), \(I\), \(\omega\) and \(\Omega\) refer to the semi-major axis, eccentricity, orbital inclination, argument of pericentre and longitude of ascending node of an orbit, with subscripts 1 and 2 referring to the inner and outer orbits, respectively. \(\mathcal{G}\) refers to the gravitational constant, while $m_0$, $m_1$ and $m_2$ are the masses of the star A, B and planet, respectively.
$\cos i$ and $\cos \alpha$ are direction cosines, where the angle $i$ corresponds to the true mutual inclination between the two orbital planes.

Then, we can compute the precession of the percientre 
of the inner orbit using 
 the Lagrange Planetary Equations \citep[e.g.][]{murray_solar_1999} as
\begin{equation}\label{eq:wdot}
    \frac{d \omega_1}{dt} = - \frac{(1-e_1^2)}{e_1G_1}\frac{\partial \mathcal{H}}{\partial e_1} + \frac{\cot I_1}{G_1}\frac{\partial \mathcal{H}}{\partial I_1},
\end{equation}
where \(G_1\) is the norm of the orbital angular momentum
\begin{equation}
    G_1 = \frac{m_0m_1}{m_0+m_1}\sqrt{\mathcal{G}(m_0 + m_1)a_1(1-e_1^2)},
\end{equation}
\begin{equation}
    \frac{\partial\mathcal{H}}{\partial e_1} = C_2\left[ - 24e_1 + 12e_1\cos^2 i + 60e_1\cos^2 \alpha \right],
\end{equation}

\begin{equation}
    \frac{\partial\mathcal{H}}{\partial I_1} = C_2\left[-12(1-e1^2)\cos i\frac{\partial\cos i}{\partial I_1}+ 60e_1^2\cos \alpha\frac{\partial\cos \alpha}{\partial I_1}\right],
\end{equation}
and 
\begin{equation}
    \frac{\partial \cos i}{\partial I_1} = \cos I_1 \sin I_2 \cos(\Omega_1-\Omega_2) - \sin I_1 \cos I_2,
\end{equation}
\begin{equation}
    \frac{\partial \cos \alpha}{\partial I_1} = \cos I_1 \sin \omega_1 \cos I_2 + \sin I_2 \sin I_1 \sin \omega_1 \cos(\Omega_1-\Omega_2) .
\end{equation}

In the case of an eclipsing binary and a transiting planet, such as Kepler-16 we can take \(I_1\approx I_2 \approx 90^\circ\), which allow us to simplify expression (\ref{eq:wdot}) as
\begin{equation}\label{eq:wdottransit}
    \begin{aligned}
     \frac{d \omega_1}{dt} & \approx \frac{12 C_2}{G_1} (1-e_1^2) \left[  2 - \cos^2 i - 5\cos^2 \alpha \right] \\
    & \approx \frac{12 C_2}{G_1} (1-e_1^2) \left[ 1 + (1 -5 \cos^2 \omega_1) \sin^2 i \right].
    \end{aligned}
\end{equation}

Therefore, for an eclipsing and transiting system, a constraint on the precession rate can be used to measure the mutual inclination.

For close-in binaries, additional sources of apsidal precession may become relevant, such as general relativity, rotational flattening and tidal deformation.
These effects, can also be modeled using a Hamiltonian formalism  as\footnote{ We assume that the spin axes of both stars are normal to the orbit.} \citep[e.g.][]{correia_tidal_2013, correia_secular_2016}
\begin{equation}
\begin{split}
    \mathcal{H}' = & - \frac{C_{g}}{(1-e_1^2)^{1/2}} - \frac{C_{r,0} + C_{r,1}}{(1-e_1^2)^{3/2}} \\ 
    & - \frac{C_{t,0} + C_{t,1}}{(1-e_1^2)^{9/2}} \left( 1 + 3 e_1^2 + \tfrac38 e_1^4 \right) \ ,
\end{split}
\end{equation}
where
\begin{equation}
C_{g} = \frac{3 \mathcal{G}^2 m_0 m_1 (m_0+m_1)}{a_1^2 c^2}
\end{equation}
corresponds to the general relativity correction ($c$ is the speed-of-the-light),
\begin{equation}
C_{r,i} = \frac{\mathcal{G} m_0 m_1 J_{2,i} R_i^2}{2 a_1^3}
\end{equation}
accounts for the rotational flattening, and
\begin{equation}
C_{t,i} = k_{2,i} \frac{\mathcal{G} m_{1-i}^2 R_i^5}{2 a_1^6}
\end{equation}
for the tidal contribution.
\begin{equation}
J_{2,i} = k_{2,i} \frac{\Omega_i^2 R_i^3}{3 \mathcal{G} m_i} \ ,
\end{equation}
$k_{2,i}$ is the second Love number for potential, $\Omega_i$ is the rotation rate, and $R_i$ is the radius of the star with mass $m_i$.

Then, according to expression (\ref{eq:wdot}), the correction in the apsidal precession is given by 
\begin{equation}
\label{eq:wdot_corr}
\begin{split}
\frac{d \omega_1}{dt} = & \; \frac{C_g}{G_1 (1-e_1^2)^{1/2}} + \frac{3 (C_{r,0} + C_{r,1})}{G_1 (1-e_1^2)^{3/2}} \\ &
+ \frac{15 (C_{t,0} + C_{t,1})}{G_1 (1-e_1^2)^{9/2}} \left( 1 + \tfrac32 e^2 + \tfrac18 e^4 \right) \ .
\end{split} 
\end{equation}
 The first term is the relativistic contribution, the middle one is the rotational contribution, and the last term is the tidal contribution. These can be taken separately as we do in Sect.~(\ref{sec:results})

\section{Tables of parameters}\label{sec:appB}
Here we show the parameters for KOI-126 in Table \ref{tab:KOI-126} and for Kepler-16 in Table \ref{tab:Kepler-16} from the fits using the binaries model. The parameters for KOI-16 are of the outer orbit of the triple, modelling the short period binary as a single massive body.

\begin{table}
	\centering
	\caption{The fitted and derived and assumed parameters for KOI-126 for the orbit of the binary B+C around star A. * since we only fit a single orbit in the radial velocity data the mass here is the combined masses of stars B and C. The quality of fit indicators (RMS, Jitter and \(\chi^2_\nu\)) are taken for the best fitting model. The mass \( M_{\rm A}\) is obtained from \citet{yenawine_photodynamical_2022}. Excepting parameters with highly asymmetric distributions, the 1\(\sigma\) uncertainties are shown as the last few significant digits.}
 	\label{tab:KOI-126}
	\begin{tabular}{lcr}
		\hline
		  & {\tt kima-binaries} & units\\
		\hline
		{\it assumed parameters} & & \\
		\( M_{\rm A}\) & 1.2713(47) & \(\mathrm{M_{\odot}}\)\\
		\hline
		{\it fitted parameters} & & \\
		\(P_{\rm obs}\)  & 33.77943(33) & days\\
		\(P_{\rm ano}\)  & 33.83207(69) & days \\
		\(K\)  & 21395(25) & \(\mathrm{m\,s^{-1}}\)\\
		\(e\)  & 0.3113(13) & \\
		\(\omega\)  & 1.1794(50) & rad\\
		\(\phi_{0}\)  & 1.0210(40) & rad\\
		\(\dot{\omega}\) & 21800(370) & \(\mathrm{arcsec\,yr^{-1}}\)\\
		\(\gamma\) & \(-27852^{+193}_{-76}\) & \(\mathrm{m\,s^{-1}}\) \\
		\hline
		{\it derived parameters} & & \\
		\(M_{\rm B+C}^{*}\)  & 0.4424(11) & \(\mathrm{M_{\odot}}\)\\
		\(T_{0}\)  & 51047.4547(27) & BJD - 2,400,000\\
		\hline
		{\it fit indicators} & & \\
		\(\rm RMS_{Tull}\) & 207 & \(\mathrm{m\,s^{-1}}\)\\
		\(\rm RMS_{Tres}\) & 84.2 & \(\mathrm{m\,s^{-1}}\)\\
		\(\rm Jitter_{Tull}\) & 0.36 & \(\mathrm{m\,s^{-1}}\)\\
		\(\rm Jitter_{Tres}\) & 49.3 & \(\mathrm{m\,s^{-1}}\)\\
		\(\chi_{\nu}^2\) & 3.07 & \\
		\hline
	\end{tabular}
\end{table}

\begin{table}
	\centering
	\caption{The fitted and derived and assumed parameters for Kepler-16 subscripts B refer to the binary orbit and b to the planetary orbit. The quality of fit indicators (RMS, Jitter and \(\chi^2_\nu\)) are taken for the best fitting model. Excepting parameters with highly asymmetric distributions, the 1\(\sigma\) uncertainties are shown as the last few significant digits.}
	\label{tab:Kepler-16}
	\begin{tabular}{lcr}
		\hline
		  & {\tt kima-binaries} & units\\
		\hline
		{\it assumed parameters} & & \\
		\( M_{\rm A}\) & 0.654(20) & \(\mathrm{M_{\odot}}\)\\
		\hline
		{\it fitted parameters} & & \\
		\(P_{\rm B,obs}\)  & 41.077716(55) & days\\
		\(P_{\rm B,ano}\)  & 41.07874(31) & days \\
		\(K_{\rm B}\)  & 13678.9(1.4) & \(\mathrm{m\,s^{-1}}\)\\
		\(e_{\rm B}\)  & 0.159925(88) & \\
		\(\omega_{\rm B}\)  & 4.60203(79) & rad\\
		\(\phi_{0, {\rm B}}\)  & 1.63340(76) & rad\\
		\(\dot{\omega}_{\rm B}\) & 284(86) & \(\mathrm{arcsec\,yr^{-1}}\)\\
		& & \\
		\(P_{\rm b}\)  & 225.8(1.7) & days\\
		\(K_{\rm b}\)  & 11.7(1.6) & \(\mathrm{m\,s^{-1}}\)\\
		\(e_{\rm b}\)  & <0.29 & \\
		\(\omega_{\rm b}\)  & 3.90(92) & rad\\
		\(\phi_{0, {\rm b}}\)  & 2.32(87) & rad\\
		& & \\
		\(\gamma\) & \(-33811.5^{+3.1}_{-0.2}\) & \(\mathrm{m\,s^{-1}}\) \\
		\hline
		{\it derived parameters} & & \\
		\(M_{B}\)  & 0.1965(32) & \(\mathrm{M_{\odot}}\)\\
		\(T_{0, {\rm B}}\)  & 58498.4796(52) & BJD - 2,400,000\\
		& & \\
		\(M_{b}\)  & 0.308(42) & \(\mathrm{M_{Jup}}\)\\
		\(T_{0, {\rm b}}\)  & 58388(33) & BJD - 2,400,000\\
		\hline
		{\it fit indicators} & & \\
		\(\rm RMS\) & 11.01 & \(\mathrm{m\,s^{-1}}\)\\
		\(\rm Jitter\) & 0.91 & \(\mathrm{m\,s^{-1}}\)\\
		\(\chi_{\nu}^2\) & 0.92 & \\
		\hline
	\end{tabular}
\end{table}
\

\bsp	
\label{lastpage}
\end{document}